\documentclass{aa} 

\usepackage{graphicx}
\usepackage{txfonts}

\usepackage[colorlinks=true,linkcolor=blue,citecolor=blue,urlcolor=blue]{hyperref}
\usepackage{orcidlink}

\def\specchar#1{{\sc{#1}}}  

\def\Heir{\mbox{He\,\specchar{i}\, 10830\, \AA}\xspace}

\begin{document}

\title{On the center-to-limb variations of the \ion{He}{I} 10\,830~\AA\ triplet}

\author{A.G.M.\ Pietrow\inst{1}\orcidlink{0000-0002-0484-7634}  
    \and C.\ Kuckein\inst{2,3}\orcidlink{0000-0002-3242-1497} 
    \and M.\ Verma\inst{1}\orcidlink{0000-0003-1054-766X}
    \and C.\ Denker\inst{1}\orcidlink{0000-0002-7729-6415}
    \and
    J.C.\ Trelles Arjona\inst{2,3}\orcidlink{0000-0001-9857-2573}
    \and R.\ Kamlah\inst{1}\orcidlink{0000-0003-2059-585X}
    \and K. Poppenhäger\inst{1}\orcidlink{0000-0003-1231-2194}}

\institute{%
    \inst{1}Leibniz-Institut für Astrophysik Potsdam (AIP), An der Sternwarte 16, 14482 Potsdam, Germany\\
    \inst{2}Instituto de Astrofísica de Canarias (IAC), Vía Láctea s/n, E-38205 La Laguna, Tenerife, Spain\\
    \inst{3}Universidad de La Laguna, Departamento de Astrofísica, E-38206 La Laguna, Tenerife, Spain\\
    \\
    \email{apietrow@aip.de}}

\date{Draft: compiled on \today}

\abstract{%
We present high-resolution spectroscopic observations of the quiet-Sun center-to-limb variations (CLV) of the \ion{He}{I} triplet at 10\,830~\AA\ and the nearby \ion{Si}{I}~10\,827~\AA\ line, observed with GREGOR Infrared Spectrograph (GRIS) and the improved High-resolution Fast Imager (HiFI+). The observations cover the interval $\mu = [0.1,\, 1.0]$, where $\mu$ is the cosine of the heliocentric angle. At each $\mu$ position, the spectra were spatially averaged over 0.02 $\mu$ and the resulting CLVs were given both as these averaged data points and as smooth polynomial curves fitted across each wavelength point.
The \ion{He}{I} spectra were inverted using the HAnle and ZEeman Light (HAZEL) code, showing an increase in optical depth towards the limb and a reversed convective blueshift for the red component, while the blue component was entirely absent. In addition, we found a strong increase in the steepness of the \ion{He}{I} CLV compared to that of the nearby continuum. The \ion{Si}{I} showed a behavior more typical of photospheric lines, namely shallower CLV, a reduction in width and depth, and a more typical convective blueshift. }


\maketitle
%

\section{Introduction}

The \ion{He}{I} triplet at 10\,830~\AA\ is a set of optically thin chromospheric lines that are sensitive to the overlying corona. They arise from a transition from the long-lived (metastable) 2$^3$S state, which is part of the triplet ($S = 1$) system of neutral helium. This is because the transition to the energetically lower singlet state is forbidden, effectively trapping the population in the triplet level \citep{Lagg2007,Leenaarts2016, Libbrecht2021}. There are currently three processes that are known to populate these levels: (1) thermal collisional excitation; (2) photoionization-recombination from \ion{He}{II} under UV illumination; and (3) nonthermal excitation from electron beams \citep[e.g.,][]{Goldberg1939,Kerr2021,Leenaarts2025}. 

Two of the triplet lines are heavily blended, with their cores forming roughly 0.1~\AA\ apart, which we refer to as  the red component. The remaining line, called the blue component, forms further away at 10829~\AA, but has a much smaller line depth \citep[e.g.,][]{Kaifler2022}.

The \ion{He}{I} triplet is uniquely sensitive to both the Zeeman and Hanle effects, making it a strong diagnostic for both strong and weak magnetic fields \citep{Bueno2002,Kuckein2009,Bethge2011}. Combined with its strong coupling to the overlying corona, this makes the triplet a valuable probe of coronal influences on the chromosphere, including the effects of coronal holes and flares \citep{Malanushenko2001, Xu2022, Anan2018, Kuckein2025}. Unlike most chromospheric lines, the triplet generally appears in absorption in active regions, producing dark plages and flares that remain in absorption for the majority of their lifetime, although transient brightenings are occasionally observed \citep[e.g.,][]{Libbrecht2019, DeWilde2025, Marena2025}.

The \ion{He}{I} 10\,830~\AA\ line is also a popular tool for detecting evaporation in exoplanet atmospheres, as its abundance makes it relatively easy to observe compared to other elements. Moreover, in contrast to Ly$\alpha$, it does not suffer from extinction by the interstellar medium or contamination from geocoronal emission \citep{Seager20000, Oklopcic2018, Spake2018, Cauley2018}. However, the aforementioned sensitivity to stellar activity does strongly influence the signal of the \ion{He}{I} 10\,830~\AA\ line in exoplanet atmospheres \citep{Poppenhaeger2022,SanzForcada2025}, meaning that rapidly evolving features could bias exoplanet transit observations.

In this work, we provide a detailed set of the quiet-Sun CLV of the \ion{He}{I} triplet and its surroundings. This will enable subsequent modeling studies to benchmark the accuracy of their models against our CLV, as is typically done for other lines \citep[e.g.,][]{unsold55, vernazza76, johan18, Ballester21}. Other studies use such profiles as initial boundary conditions to construct radiative transfer models of chromospheric and coronal structures \citep[e.g.,][]{gunar20, gunnar21, Rachmeler22} and to infer the formation height of the lines \citep{Wittmann76}. The CLV of spectral lines is also a popular tool for abundance studies \citep[e.g.,][]{Tiago2009, 2021MNRAS.508.2236B, Pietrow2023b} and a crucial input for pseudo-Sun-as-a-star studies\citep{Pietrow2024, DeWilde2025}, where quiet disks are built around observations with a limited field-of-view (FOV).

\begin{figure}[h]
\centering
\includegraphics[width=0.36\textwidth,trim={0.8cm 0.3cm 0.8cm 0.4cm},clip]{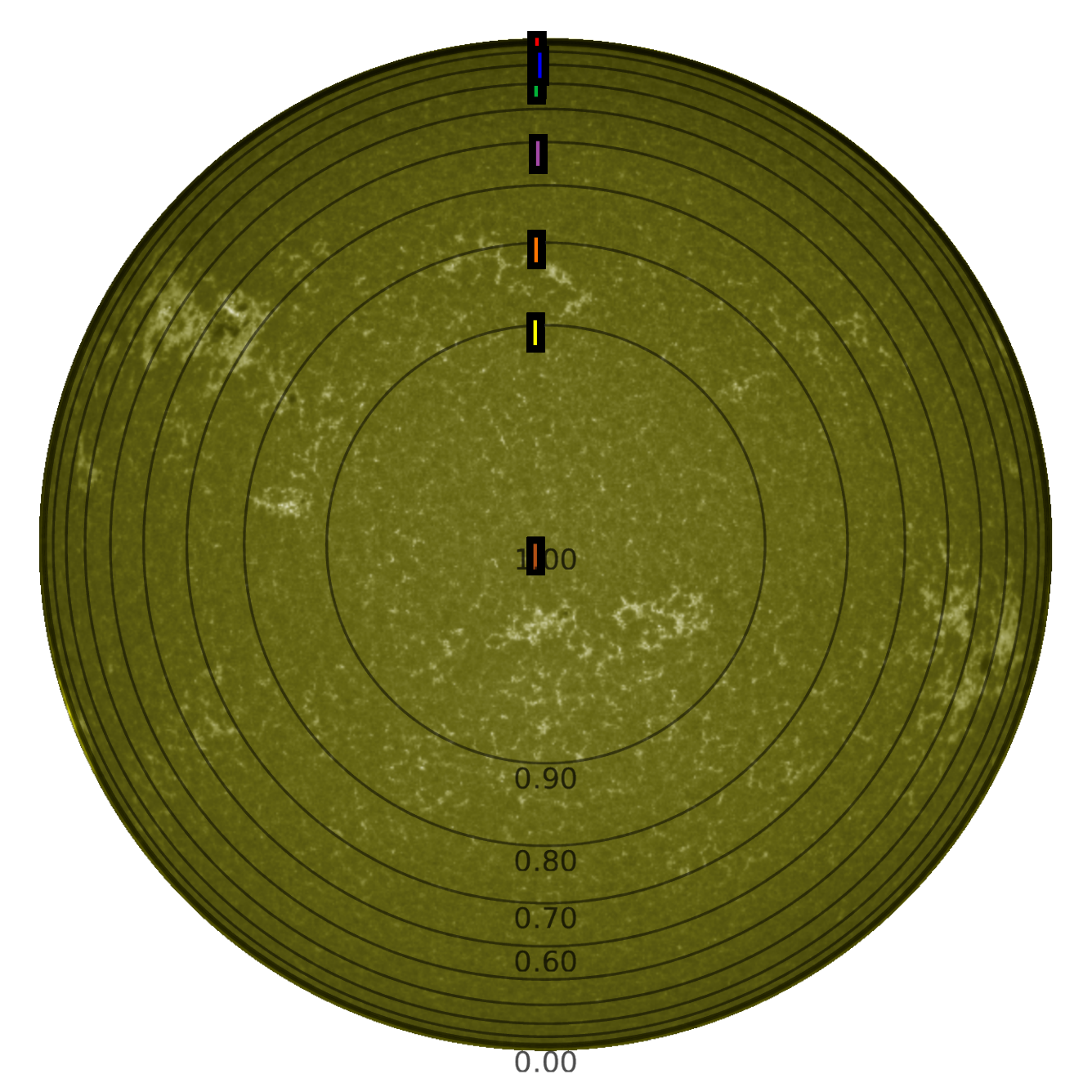}
\caption{Solar disk with the seven pointings used in this work. Ten concentric rings from $\mu=0.9$ to $\mu=0.0$ indicate steps in heliocentric angle. The background shows an AIA 1600~\AA\ image \citep[][]{Lemen2012} taken at 08:10:38~UT on the same day as the observations.}
\label{fig:pig}
\end{figure}

\section{Observations and data processing}\label{observations}

The observations were acquired with the 1.5-meter GREGOR solar telescope \citep{Schmidt2012, Kleint2020} using the improved High-resolution Fast Imager \citep[HIFI+;][]{Denker2023} and the GREGOR Infrared Spectrograph \citep[GRIS;][]{Collados2012, Regalado2024}. The data consist of a sparse mosaic spanning one solar radius, taken between 08:03~UT and 09:09~UT on 27~May 2024 (see Fig.~\ref{fig:pig} and Table~\ref{tab:1}). Each pointing consisted of 40 steps with a 0.5\arcsec\ step size of the 60\arcsec\ slit, which is sampled at 0.135\arcsec\ intervals.

For the GRIS data, a careful wavelength calibration was performed to accurately track the Doppler shifts of the faint triplet in the quiet Sun. The spectral sampling of 14.678~m\AA\ pixel$^{-1}$ was obtained from the data reduction pipeline of GRIS \citep{collados99, collados03}, which uses an average quiet-Sun profile at disk center obtained from the flat field data and matches it to the \citet{Livingston1991} Fourier Transform Spectrometers (FTS) atlas. The telluric line at 10\,832.108~\AA\ was then used to determine the zero point of the wavelength scale. Finally, the wavelength array was corrected for the gravitational redshift, the rotation of the Earth and the Sun, and the orbital motion of the Earth around the Sun. More details about these wavelength calibration steps are given in the appendices of \citet{martinez97} and \citet{kuckein12b}. This calibration resulted in a spectral window covering the wavelength range from 10\,818.50~\AA\ to 10\,833.31~\AA\ at a spectral resolution of R$\sim$190 000 (see Figs.~\ref{fig:Figinv}, ~\ref{fig:Fig1}, and~\ref{fig:Fig2}). 

The spectral line-spread function (LSF) characterizes the instrumental spectral resolution of GRIS, while the spectral stray light (or veil) represents a wavelength-independent constant contribution added to the measured intensity. Both quantities were determined by comparing an average quiet-Sun spectrum from the GRIS data with the FTS atlas, following Sect.~3 of \citet{Borrero2016A&A596A2}. The FTS spectrum was convolved with a Gaussian profile and combined with a constant veil term in the form of
\begin{equation}
I_{\mathrm{qs}}^{\mathrm{sim}}(\lambda, \sigma, \nu)
= (1 - \nu)\, I_{\mathrm{fts}}(\lambda) * g(\lambda, \sigma)
+ \nu\, I_{c,\mathrm{fts}}.\label{eq1}
\end{equation}
Here, $I^{sim}_{qs}$ is the GRIS-simulated average quiet-Sun intensity profile, $I_{fts}$ the atlas profile, $g$ a Gaussian function with width $\sigma$, $\nu$ is the fraction of spectral scattered light, and $I_{c,\mathrm{fts}}$  is the continuum intensity of the atlas close to the line.

The best agreement was obtained by minimizing the $\chi^{2}$ between the observed and synthetic spectra ($I_{\mathrm{qs}}^{\mathrm{sim}}(\lambda, \sigma, \nu)$). The resulting parameters are a Gaussian width of $\sigma = 3.77$~pixels (or $55.34$~m\AA), which corresponds to a FWHM of $130.04$~m\AA, and a spectral stray light fraction of $3.8\,\%$.

For the HiFI+ data, only the G-band channel was used since it was the longest wavelength channel available with photospheric information close to that of the Helioseismic and Magnetic Imager \citep[HMI;][]{Scherrer2012} on board the Solar Dynamics Observatory \citep[SDO;][]{Pesnell2012}. These data were processed with the sTools software \citep{Kuckein2017}, using the Kiepenheuer Institute Speckle Imaging Package \citep[KISIP;][]{Woger2008}. The speckle-restored time series covers a FOV of $71\arcsec \times 60\arcsec$ with a plate scale of 0.028\arcsec\ pixel$^{-1}$.

\section{Methods}
In this section, we discuss the data processing steps to create the CLV curves. We also present the steps we used to invert the curves.
\subsection{CLV curve generation}
Due to its high spatial resolution, the HiFI+ data were used as context information and the first frame of each observation was used to align the GREGOR FOV with respect to the HMI data. Typically, such alignments are made with a cross-correlation algorithm \citep[e.g.,][]{reardon2012, hammerschlag2013}. However, they have been shown not to work on quiet-Sun regions close to the limb \citep{Adur2020, Pietrow2023}. Instead, a manual approach is used instead, for instance, with the mosaic alignment tools from the ISPy library \citep{ISPy2021}, where bright points are used for reference. Based on the jitter of the limb, we estimated our alignment accuracy to be about two GRIS pixels, or 0.27\arcsec, which is negligible everywhere except at the extreme limb.

Once aligned to HMI, the target frames of both HiFI+ and GRIS were aligned to obtain an accurate pointing for the GRIS frames. Afterwards, a $\mu$-value was assigned to each spectrum, and the array was sorted. In \citet{Pietrow2023}, the intensity calibration was carried out by fitting a polynomial between three disk center observations, which were taken at the start, middle, and end of the mosaic to compensate for atmospheric extinction changes throughout the observing period. However, this approach could not be used, as only two disk center measurements were taken at either ends of the observations. Therefore, a new approach was taken where a continuum point, in this case 10\,823~\AA, was chosen and each $\mu$-position was set to match the predicted intensity of \citet{Neckel94} at that wavelength as the rest of the spectrum is scaled accordingly. This approach is not valid beyond $\mu = 0.08,$ as these values must be extrapolated from the rest of the fit \citep[Fig.~6 in][]{Pietrow2023}. For this reason, we did not fit our CLV beyond $\mu = 0.1$.

We average the spectra over windows of 0.02 $\mu$ for each of 10 mu positions, ranging from 1.0 to 0.1 in steps of 0.1. The number of pixels per bin over which we average is shown in Table~\ref{tab:mubin_counts}. The first bin is limited because about half of the raster was contaminated with a small filament close to the disk center. The resulting CLV is shown as a function of radius in the first panel of Fig.~\ref{fig:Fig1} and as a function of $\mu$ in the second panel. Following the method introduced by \citet{Neckel94}, the third panel shows the same values, but smoothed using a fourth-order polynomial fit across $\mu$ to each wavelength point. This smoothes the local intensity variations that are likely caused by small brightenings. This was especially helpful in the \Heir line core (see red line in the right panel of Fig.~\ref{fig:Fig5}). The resulting averaged spectra exhibited small intensity oscillations over the wavelength axis. These are likely of instrumental origin and were smoothened out using a 5-pixel (90 m\AA) box filter, which lowers the spectral resolution to $R \approx 120\,000$. For the purposes of this work, we only considered the Stokes-$I$ profiles, since the other Stokes profiles are too weak in the quiet Sun. 

\begin{table}[t]
\begingroup
\centering
\caption{GRIS raster pointings.}
\begin{tabular}{cccccc}
\hline
Pointing & Time & $X$ & $Y$ & $\mu_\mathrm{min}$ & $\mu_\mathrm{max}$\rule[-6pt]{0pt}{18pt}\\
\hline
 1 & 08:03:04~UT & $-17.1$\arcsec &   921.1\arcsec &    -- & 0.339\rule{0pt}{12pt}\\
 2 & 08:13:20~UT & $-15.6$\arcsec &   883.2\arcsec & 0.262 & 0.434\\
 3 & 08:22:11~UT & $-18.3$\arcsec &   873.7\arcsec & 0.296 & 0.455\\
 4 & 08:30:32~UT & $-15.2$\arcsec &   742.7\arcsec & 0.577 & 0.659\\
 5 & 08:46:04~UT & $-18.2$\arcsec &   562.6\arcsec & 0.779 & 0.827\\
 6 & 08:53:09~UT & $-19.6$\arcsec &   406.2\arcsec & 0.887 & 0.918\\
 7 & 09:01:31~UT & $-20.0$\arcsec & $-14.8$\arcsec & 0.998 & 1.000\rule[-6pt]{0pt}{12pt}\\
\hline
\label{tab:1}
\end{tabular}
\endgroup
\small
\noindent 
\textbf{Note:} raster pointings with the $X$- and $Y$-coordinates and the corresponding $\mu$-range for each scan. The first pointing goes off-limb.
\end{table}

\subsection{Inversions}

\begin{figure}
\centering
\includegraphics[width=1\columnwidth]{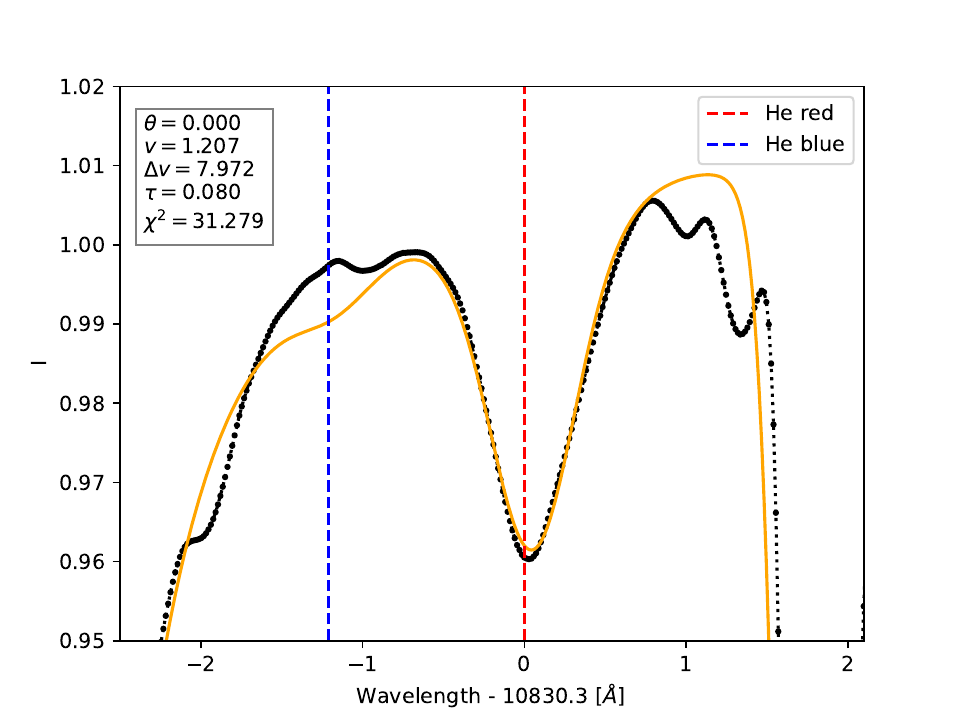}
\caption{HAZEL fit (orange) of \ion{He}{I} triplet observations at $\mu=1$ (black dashed). 
    The rest velocities of the red and blue components are marked with vertical lines in the respective colors. Basic information and the best-fit parameters are given in the box at the upper-left corner. These include the heliocentric angle, $\theta$, the Doppler velocity, $v$, the Doppler width, $\Delta v$, the optical depth, $\tau$, and the quality of fit, $\chi^2$. }
\label{fig:Figinv}
\end{figure}

We used the HAZEL code \citep{asensio08} to invert the observed spectra, considering only the Stokes-$I$ profile to retrieve a model atmosphere. The inversion strategy included a chromospheric slab model, a photospheric model, and a parametric component to account for the \ion{He}{I} triplet, the \ion{Si}{I} line at 10\,827~\AA, and the telluric line, respectively. Two inversion cycles were sufficient to achieve a good fit to the spectra (see Fig.~\ref{fig:Figinv}). 

\begin{table*}[h]
\begingroup
\centering
\caption{Number of points, $N$, per $\mu$ bin. }
\begin{tabular}{c|ccccccccccc}
\hline
$\mu$    & 1.00 & 0.90 & 0.80 & 0.70 & 0.60 & 0.50 & 0.40 & 0.30 & 0.20 & 0.10 & 0.00 \rule[-6pt]{0pt}{18pt}\\
\hline
$N$      & 17960 & 17960 & 14996 & 0 & 8423 & 0 & 9802 & 7429 & 2291 & 1126 & 59 \rule[-6pt]{0pt}{18pt}\\
\hline
\end{tabular}

\endgroup
{\small
\textbf{Note:} The $\mu=0.5$ and $\mu=0.7$ bins are empty due to the sparse sampling of the CLV (See Fig. \ref{fig:pig}).}
\label{tab:mubin_counts}
\end{table*}

\begin{figure*}
\centering
\includegraphics[width=0.99\textwidth]{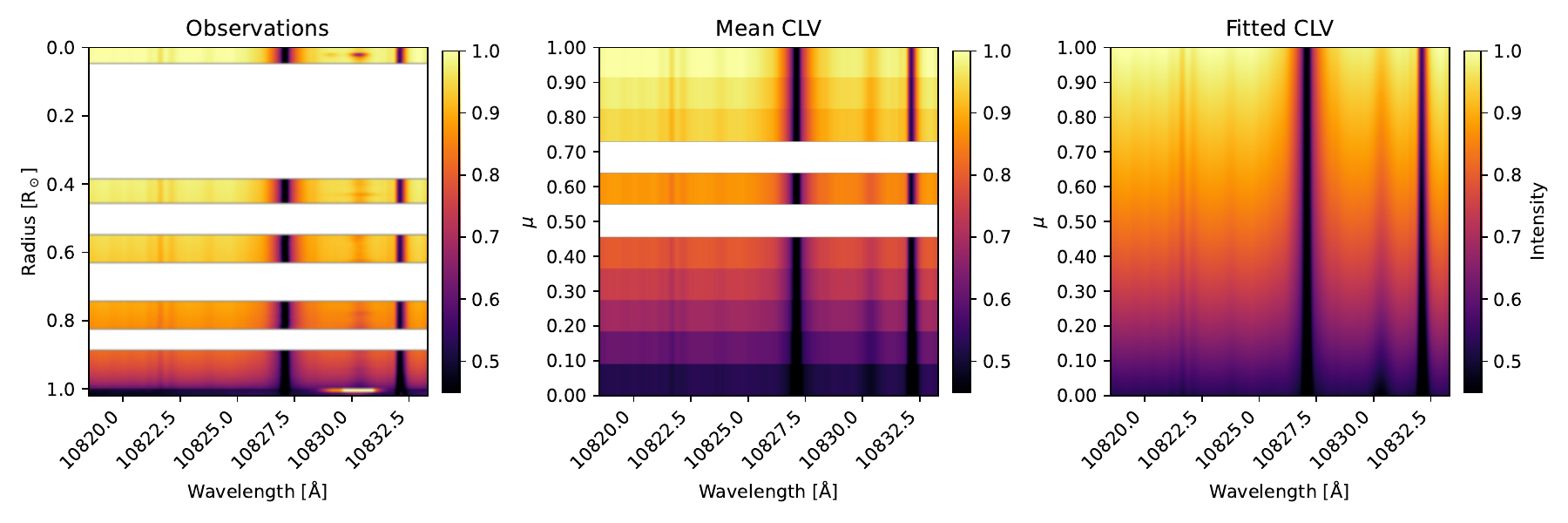}
\caption{CLV maps of the GRIS spectral window. Mean CLV profiles (\textit{left})
    across the full $\mu$-range. Oscillations are visible across the entire range. Cleaned profiles (\textit{middle}) after polynomial fitting. Normalized profiles (\textit{right}) of the middle panel, showing the relative CLV behavior across the spectral window.}
\label{fig:Fig1}
\end{figure*}

For helium, in the first cycle, only the optical depth $\tau$ and the Doppler velocity $v_\mathrm{He}$ were treated as free parameters. In the second cycle, the $v_\mathrm{He}$ was fixed and additional parameters (e.g., the Doppler width, the $\beta$ parameter, and a damping parameter) were allowed to vary. The latter two parameters primarily affect the shape of the line by modifying the depth of the core and wings. While the red component of the \ion{He}{I} line fitted well, the blue component of the \ion{He}{I} triplet was barely visible in our spectra and seems to mainly show up in active regions, such as the filament in the disk-center pointing (see first panel of Fig.~\ref{fig:Fig1}). This is in line with previous observations \citep[e.g.,][]{1994Avrett, Mauas2005, Libbrecht2019}.

For the \ion{Si}{I} line at 10\,827~\AA, the HAZEL code internally employs the Stokes Inversion based on Response functions \citep[SIR;][]{ruiz92} code, which assumes local thermodynamic equilibrium (LTE) and hydrostatic equilibrium. We used three nodes in the first cycle and five in the second for the temperature. All the other parameters were assumed to be constant with height, that is, they were assigned a single node in each cycle. Here, we only present the Doppler velocity $v_\mathrm{Si}$, as the other parameters are less reliable due to the nonLTE nature in the core of this line \citep[e.g.,][]{Bard2008}.

\begin{figure*}
\centering
\includegraphics[width=0.99\textwidth, trim={0cm 0 3.0cm 0},clip]{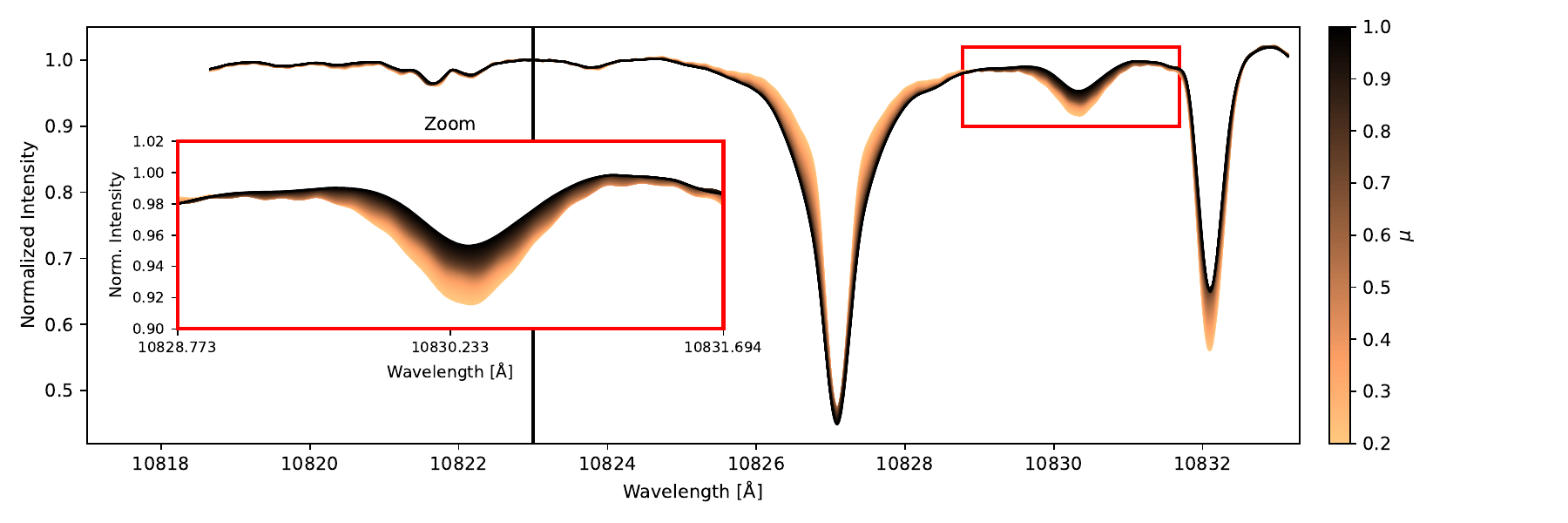}
\caption{One-dimensional version of the data in Fig.~\ref{fig:Fig1} 
    (right panel), showing the changes in the spectrum as a function of $\mu$. A zoomed-in window focuses on the \ion{He}{I} line at 10\,830~\AA. The black vertical line marks the wavelength of the intensity calibration.}
\label{fig:Fig2}
\end{figure*}

\begin{figure*}
\centering
\includegraphics[width=0.99\textwidth, trim={0cm 0 0cm 0},clip]{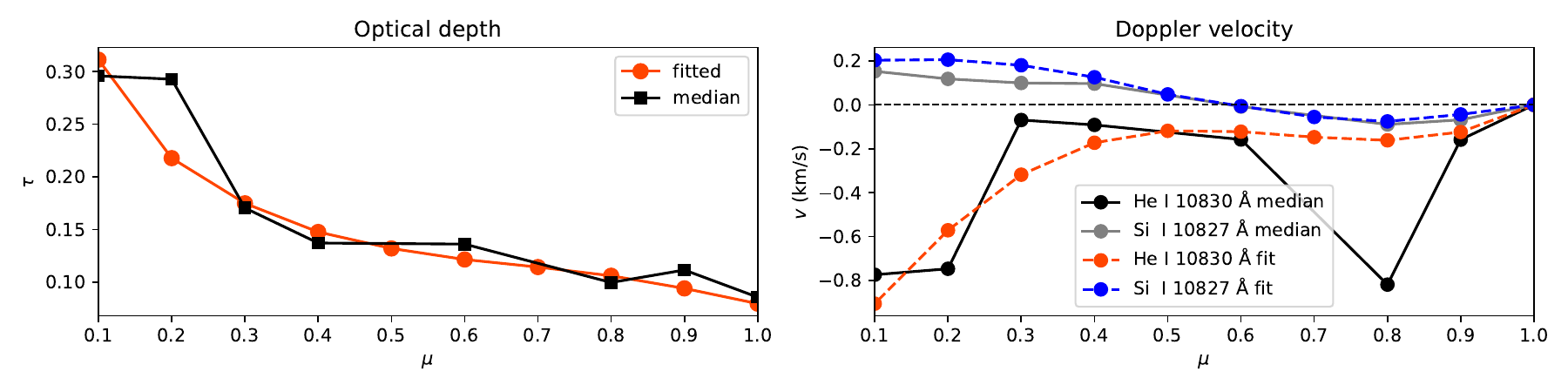}
\caption{Inferred spectral line behavior as a function of $\mu$. \textbf{Left:}  Optical depth $\tau$ of the \ion{He}{I} line. \textbf{Right:} Doppler velocity for the median \ion{He}{I} (black) and \ion{Si}{I} (gray) line profiles, together with their polynomial fits shown in red and blue dashed lines, respectively.}
\label{fig:Fig5}
\end{figure*}

\begin{figure}
\centering
\includegraphics[width=0.95\columnwidth]{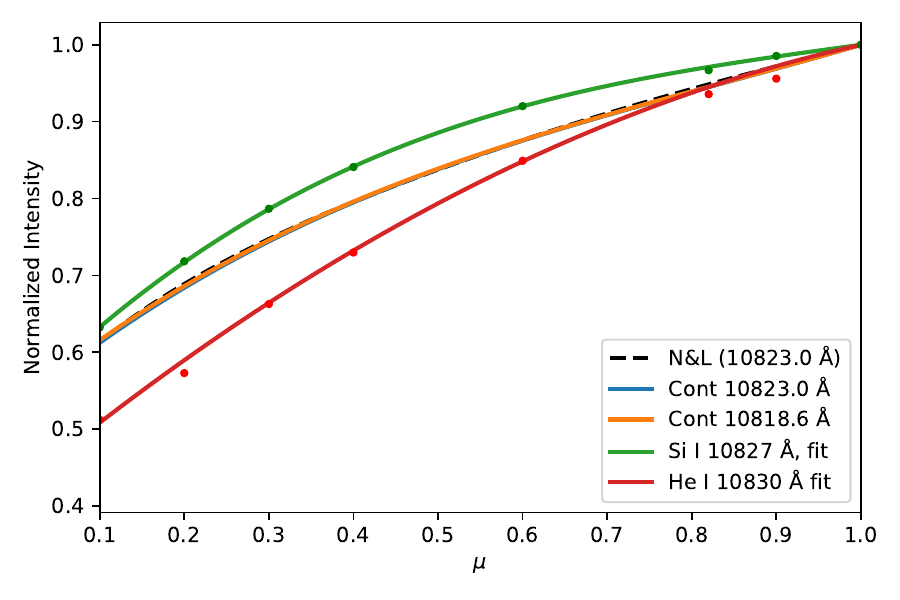}
\caption{Fitted (solid line) and mean (scattered points) CLV of the \ion{He}{I} line core at 10\,830~\AA\ (red), the
    \ion{Si}{I} line core at 10\,827~\AA\ (green), and two continuum points (blue and orange). The quiet-Sun continuum limb-darkening profile (dashed) from \citet{Neckel94} is overplotted for comparison.}
\label{fig:fig3}
\end{figure}

\section {Results and discussion}

In this section, we discuss the results inferred from both the averaged spectra and the inversions. We first discuss the \ion{Si}{I} and \Heir lines independently and then compare the results.
\subsection{\ion{Si}{I} line at 10\,827~\AA}

Inspecting Fig.~\ref{fig:Fig2}, we see that the \ion{Si}{I} line becomes slightly shallower and loses a significant part of its width when moving toward the limb. This is a typical behavior for photospheric lines, as the inclined line of sight (LOS) decreases the contribution from the lower lying layer, which contributes most of the opacity for the line wings. It also results in a shallower temperature gradient between the center and the limb, which, in turn, results in a flatter limb-darkening curve (see~Fig.~\ref{fig:fig3}). When the disk center is used as the velocity reference, such photospheric lines tend to show an apparent redshift (see the blue line in the right panel of Fig.~\ref{fig:Fig5}). This phenomenon, known as convective blueshift, arises from solar granulation. At the disk center, the observed intensity is dominated by upward-moving (blueshifted) granules, whereas towards the limb, the contribution from these granules diminishes due to the viewing geometry \citep[e.g.,][]{Lohner2019}.

\subsection{\ion{He}{I} line at 10\,830~\AA}

The \ion{He}{I} line forms in the middle to high chromosphere \citep[e.g.,][]{1994Avrett,Centeno2008}. The off-limb emission in the line peaks at around 1.5~Mm and extends to just below 3.0~Mm, which is in line with previous measurements \citep[e.g.,][]{Jaime19, Libbrecht2021}.

This line behaves quite differently from photospheric lines such as the aforementioned \ion{Si}{I} line, but also from other chromospheric lines in the visible. 
The line profile appears both deeper and broader toward the limb (Fig.~\ref{fig:Fig2}). The broadening is consistent with the behavior of other chromospheric lines reported by \citet{Pietrow2023}, although those lines become shallower with respect to the continuum because their CLV decreases less steeply than that of the nearby continuum. In contrast, the \ion{He}{I} line exhibits a comparatively steeper CLV, likely because it is optically thin and, thus, it accumulates greater opacity along the slanted LOS.

The Doppler shift relation along $\mu$ is also in the opposite direction (blue) than the aforementioned photospheric \ion{Si}{I} line  (see red+black lines and blue+gray lines in the right panel of Fig.~\ref{fig:Fig5}), resulting in a net blueshift towards the limb. The reason for this shift could be the result of an average downwards motion of the chromospheric fibrils, creating the reverse of what we see in the photosphere \citep{Wedemeyer2009, Ellwarth2023, Lohner2019},  meaning that the reason for this "fibrilar redshift" is likely rooted in the downward flows in the chromospheric canopy. 

The Doppler velocity at $\mu = 0.8$ appears as an outlier relative to the overall trend (see the right panel of Fig. \ref{fig:Fig5}). This deviation likely reflects the strong sensitivity of the \ion{He}{I} line to solar activity. While obvious filaments and bright points were removed prior to averaging, smaller features were retained to avoid biases from over-processing. Nevertheless, these outliers have little influence on the fitted polynomials, which yield a relatively smooth overall trend.

Finally, we note the complete absence of the line’s blue component in our observations. It is unlikely that this is an averaging effect similar to the one that causes the $v$ and $r$ components of the \ion{Ca}{ii}~H\&K lines to disappear \citep{Druzhinin1987}, since the blue component is also absent in the disk-center rasters outside the regions affected by filament contamination (see Fig.~\ref{fig:Fig1}). The disappearance of this component is also shown in Fig.~6 of \citet{1994Avrett}. However, it was not possible to reproduce it with HAZEL, indicating a possible unmodeled radiative transfer process. This question will be addressed further in a follow-up study.
 
\subsection{Limb darkening}
We present the CLV of the \ion{He}{I} line at 10\,830~\AA\ and the \ion{Si}{i} line at 10\,827~\AA, as well as that of the surrounding continuum in  Fig.~\ref{fig:fig3}. The values of the averaged $\mu$ positions are given as scattered points and the fitted curves as solid lines.
The continuum points are overplotted with the continuum CLV of \cite{Neckel94} to show the results of the calibration. The \ion{Si}{I} line shows a flatter curve than the continuum profile, which is expected from higher-forming photospheric lines \citep[e.g.,][]{Canocchi2024}. The \ion{He}{I} line displays a steeper curve than the surrounding continuum, a behavior that is inconsistent with other chromospheric lines. It is likely due to the line's optically thin nature, which enables greater opacity buildup along inclined lines of sight.
These results can be compared with synthetic spectra when the latter are convolved down to the smoothed resolution of GRIS following Eq.~\ref{eq1} with a $\sigma$ value of 55.34 m\AA\ and a $\nu$ value of 3.8\%.

\section{Summary and conclusions}\label{conclusions}

We present high-resolution CLV observations of the infrared \ion{He}{I} line across seven disk pointings (see Fig.~\ref{fig:pig}) obtained with GRIS at the 1.5-meter GREGOR solar telescope. These data were aligned to SDO/HMI using HiFI+ co-observations, after which a $\mu$ position was calculated for each pixel. The intensity was then calibrated by matching the continuum near 10\,824~\AA\ to the continuum limb-darkening atlas of \citet{Neckel94}. The resulting CLV spectra were then averaged into ten $\mu$ positions between 0.1 and 1.0, using an averaging window of 0.02\,$\mu$ (see Fig.~\ref{fig:Fig1}). Despite the fact that an averaging was performed over thousands of points, these bins still showed sensitivity to small levels of activity that were captured in the raster. However, these activity signals were not removed to avoid introducing additional biases. These effects were mitigated by fitting a fourth-order polynomial to each wavelength point, thus creating a smoother set of "fitted CLV" curves (see Fig.~\ref{fig:Fig2}). Low-intensity variations, likely resulting from instrumental effects, were smoothed out with a box filter.

These averaged $\mu$-bins were inverted using the HAZEL code and compared to the nearby photospheric \ion{Si}{I} line (see Fig.~\ref{fig:Fig5}). The \ion{He}{I} triplet forms under increasing opacity when closer to the limb and similarly to other chromospheric lines, it does not exhibit a convective blueshift, but, rather, a so-called fibrilar redshift. In addition, we find that the blue component is missing from our observations. The exact physical reason for this will be investigated in a follow-up work.

In Fig.~\ref{fig:fig3}, the CLVs of the \ion{He}{I} and \ion{Si}{I} lines are compared to those of the continuum and the \citet{Neckel94} continuum limb-darkening atlas. The \ion{Si}{I} line shows a flatter CLV which is characteristic of the high photosphere and chromosphere, while the \ion{He}{I} line shows a much steeper CLV. In addition, these quiet-Sun CLV measurements provide a benchmark for interpreting stellar activity signatures in this line, which can be used to support more accurate abundance and exoplanet atmosphere studies.

While a more constrained curve can likely be obtained during solar minimum, we believe that these measurements represent a significant improvement over using the continuum CLV.  These data will help us to improve the constraints on models of the \ion{He}{I} line in future studies. 

\section*{Data availability}
The CLV data is available in electronic form at the CDS via anonymous ftp to cdsarc.u-strasbg.fr (130.79.128.5) or via http://cdsweb.u-strasbg.fr/cgi-bin/qcat?J/A+A/.
\begin{acknowledgements}
We thank the anonymous referee for their valuable suggestions during the peer-review process. 

AP is supported by the Deut\-sche For\-schungs\-ge\-mein\-schaft (DFG) project number PI 2102/1-1. CK acknowledges grant RYC2022-037660-I funded by MCIN/AEI/10.13039/501100011033 and by `ESF Investing in your future'. MV acknowledges the support from IGSTC-WISER grant (IGSTC-05373). DeepL Write was used in copy editing (spelling, grammar, and readability) of the manuscript.
We extensively used the ISPy library \citep{ISPy2021} and SOAImage DS9 \citep{2003DS9} for data visualization.

\end{acknowledgements}

\bibliographystyle{aa}
\bibliography{ref}

\end{document}